\def\D       { {\mathrm{D}} } 
\def\K       { {\mathrm{K}} } 
\def\X       { {\mathrm{X}} } 
\def\simm    { {\mathrm{symm}} }

\def\DSTp    { \D^{*+} }

\def\DSTpm   { \D^{*\pm} }
\def\DCSD    { {\mathrm{DCSD}} }
\def\decDS   { \DSTp \to \D^0 \pi_{s}^+ }

\def\decDR   { \D^0 \to \K^- \pi^+ }
\def\decDW   { \D^0 \to \K^+ \pi^- }

\def\decKK   { \D^0 \to \K^- \K^+ }

\def\crm     { {\mathrm{c}} }
\def\b       { {\mathrm{b}} }
\def\e       { {\mathrm{e}} }
\def\m       { {\mathrm{m}} }
\def\cc      { \crm \bar{\crm} }
\def\bb      { \b \bar{\b} }

\def\stat    { {\mathrm{stat.}} }
\def\syst    { {\mathrm{syst.}} }
\def\pt      { p_{{\mathrm{T}}} } 
\def\MeV     { {\mathrm{MeV}} }

\def\MeVcc   { \MeV/c^2 }
\def\GeV     { {\mathrm{GeV}} }
\def\GeVc    { \GeV/c }
\def\GeVcc   { \GeV/c^2 }

\def \mx    { {\mathrm{mix}} }

\def \dirc   {dE/dx} 
\def \revc   { \overline{dE/dx} }

\newcommand{\dEdx}{\mbox{$dE/dx$}}
\newcommand{\chipi}{\mbox{$\chi_{\pi}$}}
\newcommand{\chik}{\mbox{$\chi_{\K}$}}

\documentstyle[12pt,epsfig]{article}
\begin{document} 
\begin{titlepage}

\begin{center}
{\Large EUROPEAN LABORATORY FOR PARTICLE PHYSICS}
\begin{flushright}

\vskip 1cm
\begin{tabular}{r}
CERN-EP/98-118 \\
20 July 1998 \\
\end{tabular}
\end{flushright}
 
\vskip 2cm

\begin{center}
\boldmath
\centerline{\Large\bf Study of $\D^0$--${\bar\D^0}$ Mixing and }
\centerline{\Large\bf $\D^0$ Doubly Cabibbo-Suppressed Decays}
\unboldmath
\vskip .8 cm
 
{\bf The ALEPH Collaboration$^*$}\\
\end{center}

\vskip  2.8cm
{\bf Abstract}
\end{center} 
Using a sample of four million hadronic Z events collected in ALEPH
from 1991 to 1995, the decays $\decDS$, with $\D^0$ decaying to
$\K^{-} \pi^{+}$ or to $\K^{+} \pi^{-}$, are studied. The relative
branching ratio $B(\decDW) / B(\decDR)$ is measured to be $\left( 1.84
\pm 0.59(\stat)\right.$ $\left.\pm 0.34(\syst) \right) \! \%.$ The two
possible contributions to the $\decDW$ decay, doubly
Cabibbo-suppressed decays and $\D^{0}$--$\bar{\D}^0$ mixing, are
disentangled by measuring the proper-time distribution of the
reconstructed $\D^0$'s.  Assuming no interference between the two
processes, the upper limit obtained on the mixing rate is $ 0.92 \%$
${\mathrm{at}} \ 95\% \ {\mathrm{CL}}$.  The possible effect of
interference between the two amplitudes is also assessed.

\vskip  2cm

\begin{center}
(Submitted to Physics Letters B)
\end{center}
\vfill
\vbox{
\hrule width6cm height0.5pt\vskip4pt
\noindent $^*$See the following pages for the list of authors.}

\end{titlepage}

\pagestyle{empty}
\newpage
\small
%
\newlength{\saveparskip}
\newlength{\savetextheight}
\newlength{\savetopmargin}
\newlength{\savetextwidth}
\newlength{\saveoddsidemargin}
\newlength{\savetopsep}
\setlength{\saveparskip}{\parskip}
\setlength{\savetextheight}{\textheight}
\setlength{\savetopmargin}{\topmargin}
\setlength{\savetextwidth}{\textwidth}
\setlength{\saveoddsidemargin}{\oddsidemargin}
\setlength{\savetopsep}{\topsep}
%
%
\setlength{\parskip}{0.0cm}
\setlength{\textheight}{25.0cm}
\setlength{\topmargin}{-1.5cm}
\setlength{\textwidth}{16 cm}
\setlength{\oddsidemargin}{-0.0cm}
\setlength{\topsep}{1mm}
\pretolerance=10000
\centerline{\large\bf The ALEPH Collaboration}
\footnotesize
\vspace{0.5cm}
{\raggedbottom
\begin{sloppypar}
\samepage\noindent
R.~Barate,
D.~Buskulic,
D.~Decamp,
P.~Ghez,
C.~Goy,
\mbox{J.-P.~Lees},
A.~Lucotte,
E.~Merle,
\mbox{M.-N.~Minard},
\mbox{J.-Y.~Nief},
B.~Pietrzyk
\nopagebreak
\begin{center}
\parbox{15.5cm}{\sl\samepage
Laboratoire de Physique des Particules (LAPP), IN$^{2}$P$^{3}$-CNRS,
F-74019 Annecy-le-Vieux Cedex, France}
\end{center}\end{sloppypar}
\vspace{2mm}
\begin{sloppypar}
\noindent
R.~Alemany,
G.~Boix,
M.P.~Casado,
M.~Chmeissani,
J.M.~Crespo,
M.~Delfino,
E.~Fernandez,
\mbox{M.~Fernandez-Bosman},
Ll.~Garrido,$^{15}$
E.~Graug\`{e}s,
A.~Juste,
M.~Martinez,
G.~Merino,
R.~Miquel,
Ll.M.~Mir,
I.C.~Park,
A.~Pascual,
I.~Riu,
F.~Sanchez
\nopagebreak
\begin{center}
\parbox{15.5cm}{\sl\samepage
Institut de F\'{i}sica d'Altes Energies, Universitat Aut\`{o}noma
de Barcelona, E-08193 Bellaterra (Barcelona), Spain$^{7}$}
\end{center}\end{sloppypar}
\vspace{2mm}
\begin{sloppypar}
\noindent
A.~Colaleo,
D.~Creanza,
M.~de~Palma,
G.~Gelao,
G.~Iaselli,
G.~Maggi,
M.~Maggi,
S.~Nuzzo,
A.~Ranieri,
G.~Raso,
F.~Ruggieri,
G.~Selvaggi,
L.~Silvestris,
P.~Tempesta,
A.~Tricomi,$^{3}$
G.~Zito
\nopagebreak
\begin{center}
\parbox{15.5cm}{\sl\samepage
Dipartimento di Fisica, INFN Sezione di Bari, I-70126
Bari, Italy}
\end{center}\end{sloppypar}
\vspace{2mm}
\begin{sloppypar}
\noindent
X.~Huang,
J.~Lin,
Q. Ouyang,
T.~Wang,
Y.~Xie,
R.~Xu,
S.~Xue,
J.~Zhang,
L.~Zhang,
W.~Zhao
\nopagebreak
\begin{center}
\parbox{15.5cm}{\sl\samepage
Institute of High-Energy Physics, Academia Sinica, Beijing, The People's
Republic of China$^{8}$}
\end{center}\end{sloppypar}
\vspace{2mm}
\begin{sloppypar}
\noindent
D.~Abbaneo,
U.~Becker,
\mbox{P.~Bright-Thomas},$^{24}$
D.~Casper,
M.~Cattaneo,
F.~Cerutti,
V.~Ciulli,
G.~Dissertori,
H.~Drevermann,
R.W.~Forty,
M.~Frank,
R.~Hagelberg,
A.W. Halley,
J.B.~Hansen,
J.~Harvey,
P.~Janot,
B.~Jost,
I.~Lehraus,
P.~Mato,
A.~Minten,
L.~Moneta,$^{21}$
A.~Pacheco,
F.~Ranjard,
L.~Rolandi,
D.~Rousseau,
D.~Schlatter,
M.~Schmitt,$^{20}$
O.~Schneider,
W.~Tejessy,
F.~Teubert,
I.R.~Tomalin,
H.~Wachsmuth
\nopagebreak
\begin{center}
\parbox{15.5cm}{\sl\samepage
European Laboratory for Particle Physics (CERN), CH-1211 Geneva 23,
Switzerland}
\end{center}\end{sloppypar}
\vspace{2mm}
\begin{sloppypar}
\noindent
Z.~Ajaltouni,
F.~Badaud,
G.~Chazelle,
O.~Deschamps,
A.~Falvard,
C.~Ferdi,
P.~Gay,
C.~Guicheney,
P.~Henrard,
J.~Jousset,
B.~Michel,
S.~Monteil,
\mbox{J-C.~Montret},
D.~Pallin,
P.~Perret,
F.~Podlyski,
J.~Proriol,
P.~Rosnet
\nopagebreak
\begin{center}
\parbox{15.5cm}{\sl\samepage
Laboratoire de Physique Corpusculaire, Universit\'e Blaise Pascal,
IN$^{2}$P$^{3}$-CNRS, Clermont-Ferrand, F-63177 Aubi\`{e}re, France}
\end{center}\end{sloppypar}
\vspace{2mm}
\begin{sloppypar}
\noindent
J.D.~Hansen,
J.R.~Hansen,
P.H.~Hansen,
B.S.~Nilsson,
B.~Rensch,
A.~W\"a\"an\"anen
\begin{center}
\parbox{15.5cm}{\sl\samepage
Niels Bohr Institute, DK-2100 Copenhagen, Denmark$^{9}$}
\end{center}\end{sloppypar}
\vspace{2mm}
\begin{sloppypar}
\noindent
G.~Daskalakis,
A.~Kyriakis,
C.~Markou,
E.~Simopoulou,
I.~Siotis,
A.~Vayaki
\nopagebreak
\begin{center}
\parbox{15.5cm}{\sl\samepage
Nuclear Research Center Demokritos (NRCD), GR-15310 Attiki, Greece}
\end{center}\end{sloppypar}
\vspace{2mm}
\begin{sloppypar}
\noindent
A.~Blondel,
G.~Bonneaud,
\mbox{J.-C.~Brient},
P.~Bourdon,
A.~Roug\'{e},
M.~Rumpf,
A.~Valassi,$^{6}$
M.~Verderi,
H.~Videau
\nopagebreak
\begin{center}
\parbox{15.5cm}{\sl\samepage
Laboratoire de Physique Nucl\'eaire et des Hautes Energies, Ecole
Polytechnique, IN$^{2}$P$^{3}$-CNRS, \mbox{F-91128} Palaiseau Cedex, France}
\end{center}\end{sloppypar}
\vspace{2mm}
\begin{sloppypar}
\noindent
E.~Focardi,
G.~Parrini,
K.~Zachariadou
\nopagebreak
\begin{center}
\parbox{15.5cm}{\sl\samepage
Dipartimento di Fisica, Universit\`a di Firenze, INFN Sezione di Firenze,
I-50125 Firenze, Italy}
\end{center}\end{sloppypar}
\vspace{2mm}
\begin{sloppypar}
\noindent
M.~Corden,
C.~Georgiopoulos,
D.E.~Jaffe
\nopagebreak
\begin{center}
\parbox{15.5cm}{\sl\samepage
Supercomputer Computations Research Institute,
Florida State University,
Tallahassee, FL 32306-4052, USA $^{13,14}$}
\end{center}\end{sloppypar}
\vspace{2mm}
\begin{sloppypar}
\noindent
A.~Antonelli,
G.~Bencivenni,
G.~Bologna,$^{4}$
F.~Bossi,
P.~Campana,
G.~Capon,
V.~Chiarella,
G.~Felici,
P.~Laurelli,
G.~Mannocchi,$^{5}$
F.~Murtas,
G.P.~Murtas,
L.~Passalacqua,
\mbox{M.~Pepe-Altarelli}
\nopagebreak
\begin{center}
\parbox{15.5cm}{\sl\samepage
Laboratori Nazionali dell'INFN (LNF-INFN), I-00044 Frascati, Italy}
\end{center}\end{sloppypar}
\vspace{2mm}
\pagebreak
\begin{sloppypar}
\noindent
L.~Curtis,
J.G.~Lynch,
P.~Negus,
V.~O'Shea,
C.~Raine,
J.M.~Scarr,
K.~Smith,
\mbox{P.~Teixeira-Dias},
A.S.~Thompson,
E.~Thomson
\nopagebreak
\begin{center}
\parbox{15.5cm}{\sl\samepage
Department of Physics and Astronomy, University of Glasgow, Glasgow G12
8QQ,United Kingdom$^{10}$}
\end{center}\end{sloppypar}
\vspace{2mm}
\begin{sloppypar}
\noindent
O.~Buchm\"uller,
S.~Dhamotharan,
C.~Geweniger,
G.~Graefe,
P.~Hanke,
G.~Hansper,
V.~Hepp,
E.E.~Kluge,
A.~Putzer,
J.~Sommer,
K.~Tittel,
S.~Werner,
M.~Wunsch
\nopagebreak
\begin{center}
\parbox{15.5cm}{\sl\samepage
Institut f\"ur Hochenergiephysik, Universit\"at Heidelberg, D-69120
Heidelberg, Germany$^{16}$}
\end{center}\end{sloppypar}
\vspace{2mm}
\begin{sloppypar}
\noindent
R.~Beuselinck,
D.M.~Binnie,
W.~Cameron,
P.J.~Dornan,$^{2}$
M.~Girone,
S.~Goodsir,
E.B.~Martin,
N.~Marinelli,
A.~Moutoussi,
J.~Nash,
J.K.~Sedgbeer,
P.~Spagnolo,
M.D.~Williams
\nopagebreak
\begin{center}
\parbox{15.5cm}{\sl\samepage
Department of Physics, Imperial College, London SW7 2BZ,
United Kingdom$^{10}$}
\end{center}\end{sloppypar}
\vspace{2mm}
\begin{sloppypar}
\noindent
V.M.~Ghete,
P.~Girtler,
E.~Kneringer,
D.~Kuhn,
G.~Rudolph
\nopagebreak
\begin{center}
\parbox{15.5cm}{\sl\samepage
Institut f\"ur Experimentalphysik, Universit\"at Innsbruck, A-6020
Innsbruck, Austria$^{18}$}
\end{center}\end{sloppypar}
\vspace{2mm}
\begin{sloppypar}
\noindent
A.P.~Betteridge,
C.K.~Bowdery,
P.G.~Buck,
P.~Colrain,
G.~Crawford,
A.J.~Finch,
F.~Foster,
G.~Hughes,
R.W.L.~Jones,
N.A.~Robertson,
M.I.~Williams
\nopagebreak
\begin{center}
\parbox{15.5cm}{\sl\samepage
Department of Physics, University of Lancaster, Lancaster LA1 4YB,
United Kingdom$^{10}$}
\end{center}\end{sloppypar}
\vspace{2mm}
\begin{sloppypar}
\noindent
I.~Giehl,
C.~Hoffmann,
K.~Jakobs,
K.~Kleinknecht,
G.~Quast,
B.~Renk,
E.~Rohne,
\mbox{H.-G.~Sander},
P.~van~Gemmeren,
C.~Zeitnitz
\nopagebreak
\begin{center}
\parbox{15.5cm}{\sl\samepage
Institut f\"ur Physik, Universit\"at Mainz, D-55099 Mainz, Germany$^{16}$}
\end{center}\end{sloppypar}
\vspace{2mm}
\begin{sloppypar}
\noindent
J.J.~Aubert,
C.~Benchouk,
A.~Bonissent,
G.~Bujosa,
J.~Carr,$^{2}$
P.~Coyle,
F.~Etienne,
O.~Leroy,
F.~Motsch,
P.~Payre,
M.~Talby,
A.~Sadouki,
M.~Thulasidas,
K.~Trabelsi
\nopagebreak
\begin{center}
\parbox{15.5cm}{\sl\samepage
Centre de Physique des Particules, Facult\'e des Sciences de Luminy,
IN$^{2}$P$^{3}$-CNRS, F-13288 Marseille, France}
\end{center}\end{sloppypar}
\vspace{2mm}
\begin{sloppypar}
\noindent
M.~Aleppo,
M.~Antonelli,
F.~Ragusa
\nopagebreak
\begin{center}
\parbox{15.5cm}{\sl\samepage
Dipartimento di Fisica, Universit\`a di Milano e INFN Sezione di Milano,
I-20133 Milano, Italy}
\end{center}\end{sloppypar}
\vspace{2mm}
\begin{sloppypar}
\noindent
R.~Berlich,
V.~B\"uscher,
G.~Cowan,
H.~Dietl,
G.~Ganis,
G.~L\"utjens,
C.~Mannert,
W.~M\"anner,
\mbox{H.-G.~Moser},
S.~Schael,
R.~Settles,
H.~Seywerd,
H.~Stenzel,
W.~Wiedenmann,
G.~Wolf
\nopagebreak
\begin{center}
\parbox{15.5cm}{\sl\samepage
Max-Planck-Institut f\"ur Physik, Werner-Heisenberg-Institut,
D-80805 M\"unchen, Germany\footnotemark[16]}
\end{center}\end{sloppypar}
\vspace{2mm}
\begin{sloppypar}
\noindent
J.~Boucrot,
O.~Callot,
S.~Chen,
A.~Cordier,
M.~Davier,
L.~Duflot,
\mbox{J.-F.~Grivaz},
Ph.~Heusse,
A.~H\"ocker,
A.~Jacholkowska,
D.W.~Kim,$^{12}$
F.~Le~Diberder,
J.~Lefran\c{c}ois,
\mbox{A.-M.~Lutz},
\mbox{M.-H.~Schune},
E.~Tournefier,
\mbox{J.-J.~Veillet},
I.~Videau,
D.~Zerwas
\nopagebreak
\begin{center}
\parbox{15.5cm}{\sl\samepage
Laboratoire de l'Acc\'el\'erateur Lin\'eaire, Universit\'e de Paris-Sud,
IN$^{2}$P$^{3}$-CNRS, F-91898 Orsay Cedex, France}
\end{center}\end{sloppypar}
\vspace{2mm}
\begin{sloppypar}
\noindent
\samepage
P.~Azzurri,
G.~Bagliesi,$^{2}$
G.~Batignani,
S.~Bettarini,
T.~Boccali,
C.~Bozzi,
G.~Calderini,
M.~Carpinelli,
M.A.~Ciocci,
R.~Dell'Orso,
R.~Fantechi,
I.~Ferrante,
L.~Fo\`{a},$^{1}$
F.~Forti,
A.~Giassi,
M.A.~Giorgi,
A.~Gregorio,
F.~Ligabue,
A.~Lusiani,
P.S.~Marrocchesi,
A.~Messineo,
F.~Palla,
G.~Rizzo,
G.~Sanguinetti,
A.~Sciab\`a,
G.~Sguazzoni,
R.~Tenchini,
G.~Tonelli,$^{19}$
C.~Vannini,
A.~Venturi,
P.G.~Verdini
\samepage
\begin{center}
\parbox{15.5cm}{\sl\samepage
Dipartimento di Fisica dell'Universit\`a, INFN Sezione di Pisa,
e Scuola Normale Superiore, I-56010 Pisa, Italy}
\end{center}\end{sloppypar}
\vspace{2mm}
\begin{sloppypar}
\noindent
G.A.~Blair,
L.M.~Bryant,
J.T.~Chambers,
M.G.~Green,
T.~Medcalf,
P.~Perrodo,
J.A.~Strong,
\mbox{J.H.~von~Wimmersperg-Toeller}
\nopagebreak
\begin{center}
\parbox{15.5cm}{\sl\samepage
Department of Physics, Royal Holloway \& Bedford New College,
University of London, Surrey TW20 OEX, United Kingdom$^{10}$}
\end{center}\end{sloppypar}
\vspace{2mm}
\begin{sloppypar}
\noindent
D.R.~Botterill,
R.W.~Clifft,
T.R.~Edgecock,
P.R.~Norton,
J.C.~Thompson,
A.E.~Wright
\nopagebreak
\begin{center}
\parbox{15.5cm}{\sl\samepage
Particle Physics Dept., Rutherford Appleton Laboratory,
Chilton, Didcot, Oxon OX11 OQX, United Kingdom$^{10}$}
\end{center}\end{sloppypar}
\vspace{2mm}
\begin{sloppypar}
\noindent
\mbox{B.~Bloch-Devaux},
P.~Colas,
S.~Emery,
W.~Kozanecki,
E.~Lan\c{c}on,$^{2}$
\mbox{M.-C.~Lemaire},
E.~Locci,
P.~Perez,
J.~Rander,
\mbox{J.-F.~Renardy},
A.~Roussarie,
\mbox{J.-P.~Schuller},
J.~Schwindling,
A.~Trabelsi,
B.~Vallage
\nopagebreak
\begin{center}
\parbox{15.5cm}{\sl\samepage
CEA, DAPNIA/Service de Physique des Particules,
CE-Saclay, F-91191 Gif-sur-Yvette Cedex, France$^{17}$}
\end{center}\end{sloppypar}
\pagebreak
\vspace{2mm}
\begin{sloppypar}
\noindent
S.N.~Black,
J.H.~Dann,
R.P.~Johnson,
H.Y.~Kim,
N.~Konstantinidis,
A.M.~Litke,
M.A. McNeil,
G.~Taylor
\nopagebreak
\begin{center}
\parbox{15.5cm}{\sl\samepage
Institute for Particle Physics, University of California at
Santa Cruz, Santa Cruz, CA 95064, USA$^{22}$}
\end{center}\end{sloppypar}
\vspace{2mm}
\begin{sloppypar}
\noindent
C.N.~Booth,
S.~Cartwright,
F.~Combley,
M.S.~Kelly,
M.~Lehto,
L.F.~Thompson
\nopagebreak
\begin{center}
\parbox{15.5cm}{\sl\samepage
Department of Physics, University of Sheffield, Sheffield S3 7RH,
United Kingdom$^{10}$}
\end{center}\end{sloppypar}
\vspace{2mm}
\begin{sloppypar}
\noindent
K.~Affholderbach,
A.~B\"ohrer,
S.~Brandt,
C.~Grupen,
P.~Saraiva,
L.~Smolik,
F.~Stephan
\nopagebreak
\begin{center}
\parbox{15.5cm}{\sl\samepage
Fachbereich Physik, Universit\"at Siegen, D-57068 Siegen,
 Germany$^{16}$}
\end{center}\end{sloppypar}
\vspace{2mm}
\begin{sloppypar}
\noindent
G.~Giannini,
B.~Gobbo,
G.~Musolino
\nopagebreak
\begin{center}
\parbox{15.5cm}{\sl\samepage
Dipartimento di Fisica, Universit\`a di Trieste e INFN Sezione di Trieste,
I-34127 Trieste, Italy}
\end{center}\end{sloppypar}
\vspace{2mm}
\begin{sloppypar}
\noindent
J.~Rothberg,
S.~Wasserbaech
\nopagebreak
\begin{center}
\parbox{15.5cm}{\sl\samepage
Experimental Elementary Particle Physics, University of Washington, WA 98195
Seattle, U.S.A.}
\end{center}\end{sloppypar}
\vspace{2mm}
\begin{sloppypar}
\noindent
S.R.~Armstrong,
E.~Charles,
P.~Elmer,
D.P.S.~Ferguson,
Y.~Gao,
S.~Gonz\'{a}lez,
T.C.~Greening,
O.J.~Hayes,
H.~Hu,
S.~Jin,
P.A.~McNamara III,
J.M.~Nachtman,$^{23}$
J.~Nielsen,
W.~Orejudos,
Y.B.~Pan,
Y.~Saadi,
I.J.~Scott,
J.~Walsh,
Sau~Lan~Wu,
X.~Wu,
G.~Zobernig
\nopagebreak
\begin{center}
\parbox{15.5cm}{\sl\samepage
Department of Physics, University of Wisconsin, Madison, WI 53706,
USA$^{11}$}
\end{center}\end{sloppypar}
}
\footnotetext[1]{Now at CERN, 1211 Geneva 23,
Switzerland.}
\footnotetext[2]{Also at CERN, 1211 Geneva 23, Switzerland.}
\footnotetext[3]{Also at Dipartimento di Fisica, INFN, Sezione di Catania, Catania, Italy.}
\footnotetext[4]{Also Istituto di Fisica Generale, Universit\`{a} di
Torino, Torino, Italy.}
\footnotetext[5]{Also Istituto di Cosmo-Geofisica del C.N.R., Torino,
Italy.}
\footnotetext[6]{Supported by the Commission of the European Communities,
contract ERBCHBICT941234.}
\footnotetext[7]{Supported by CICYT, Spain.}
\footnotetext[8]{Supported by the National Science Foundation of China.}
\footnotetext[9]{Supported by the Danish Natural Science Research Council.}
\footnotetext[10]{Supported by the UK Particle Physics and Astronomy Research
Council.}
\footnotetext[11]{Supported by the US Department of Energy, grant
DE-FG0295-ER40896.}
\footnotetext[12]{Permanent address: Kangnung National University, Kangnung, 
Korea.}
\footnotetext[13]{Supported by the US Department of Energy, contract
DE-FG05-92ER40742.}
\footnotetext[14]{Supported by the US Department of Energy, contract
DE-FC05-85ER250000.}
\footnotetext[15]{Permanent address: Universitat de Barcelona, 08208 Barcelona,
Spain.}
\footnotetext[16]{Supported by the Bundesministerium f\"ur Bildung,
Wissenschaft, Forschung und Technologie, Germany.}
\footnotetext[17]{Supported by the Direction des Sciences de la
Mati\`ere, C.E.A.}
\footnotetext[18]{Supported by Fonds zur F\"orderung der wissenschaftlichen
Forschung, Austria.}
\footnotetext[19]{Also at Istituto di Matematica e Fisica,
Universit\`a di Sassari, Sassari, Italy.}
\footnotetext[20]{Now at Harvard University, Cambridge, MA 02138, U.S.A.}
\footnotetext[21]{Now at University of Geneva, 1211 Geneva 4, Switzerland.}
\footnotetext[22]{Supported by the US Department of Energy,
grant DE-FG03-92ER40689.}
\footnotetext[23]{Now at University of California at Los Angeles (UCLA),
Los Angeles, CA 90024, U.S.A.}
\footnotetext[24]{Now at School of Physics and Astronomy,
Birmingham B15 2TT, U.K.}
%
%
\setlength{\parskip}{\saveparskip}
\setlength{\textheight}{\savetextheight}
\setlength{\topmargin}{\savetopmargin}
\setlength{\textwidth}{\savetextwidth}
\setlength{\oddsidemargin}{\saveoddsidemargin}
\setlength{\topsep}{\savetopsep}
\normalsize
\newpage
\pagestyle{plain}
\setcounter{page}{1}

\pagestyle{plain}
\setcounter{page}{1}
\setcounter{footnote}{0}

\section{Introduction}

The $\D^0$ can produce a $\K^+ \pi^-$ system either via a doubly
Cabibbo-suppressed decay (DCSD) or via the oscillation of the $\D^0$
into a $\bar{\D}^0$ followed by the Cabibbo-favoured decay $\bar\D^{0}
\to \K^+ \pi^-$.  The rate of DCSD processes $\decDW$ is expected to
be of the order of $2 \tan^4\theta_C \sim 0.6 \%$~\cite{PRED}, where
$\theta_C$ is the Cabibbo angle.  Within the framework of the Standard
Model the $\D^{0}$--$\bar{\D}^0$ mixing rate $R_{\mx}$ is expected to
be well below present experimental bounds~\cite{Georgi,Burdman}. While
short distance effects from box diagrams are known to give a small
contribution ($R_{\mx} \sim 10^{-10}$)~\cite{Gaillard}, long distance
effects from second-order weak interactions with mesonic intermediate
states may give a much larger contribution but are subject to large
theoretical uncertainties ($R_{\mx} \sim
10^{-7}-10^{-3}$)~\cite{Wolfenstein}.

There are many extensions of the Standard Model which allow a
$\D^0$--$\bar{\D^0}$ mixing rate significantly larger than the Standard
Model prediction, for example models with leptoquarks, with two-Higgs-doublet, 
with fourth quark generation and supersymmetric models with
alignment~\cite{BP,NS}. Experimental evidence for mixing within the
current experimental sensitivity would therefore be an indication of new
physics.

The search for DCSD or mixing necessitates the identification of a
change in the charm quantum number between production and decay of the
$\D^0$.  The method presented here consists of reconstructing the
$\decDS$ decay where the charge of the slow pion indicates whether a
$\D^0$ or a $\bar{\D}^0$ is produced.  The charge of the kaon in the
subsequent $\D^0 \to \K \pi$ decay tags the charm flavour at decay.
The relative contributions of the DCSD process and the
$\D^{0}$--$\bar{\D}^0$ mixing are assessed by studying the proper-time
distribution of the reconstructed $\D^0$'s.

\section{ALEPH Detector and Data Selection} 

This analysis uses data collected in the vicinity of the Z peak from
1991 to 1995 with the ALEPH detector at the LEP electron-positron
storage ring.  The data sample consists of about four million hadronic
Z decays that satisfy the criteria of Ref.~\cite{Adron}.

A detailed description of the design and performance of the
apparatus can be found in Refs.~\cite{ALEPH1,ALEPH2}, and only a brief
summary of the features relevant to this study is given here. A
double-sided silicon vertex detector (VDET), surrounding the beam
pipe, is installed close to the interaction region. The single-hit
resolution for the $ r \phi$ and $z$ projections is
12~$\mu\m$. Outside the vertex detector are an eight-layer drift
chamber, the inner tracking chamber (ITC), and a large time projection
chamber (TPC). These three detectors form the tracking system, which
is immersed in a 1.5 T axial magnetic field. Using the VDET, ITC and
TPC coordinates the particle momentum transverse to the beam axis is
measured with a resolution of $\delta\pt/\pt=6 \times 10^{-4}\pt
\oplus 5\times 10^{-3}$  ($\pt$ in $\GeVc$).

The TPC also provides up to 338~measurements of the specific ionization
of a charged particle. In the following, the \dEdx\ information is
considered as available if more than 50 samples are present.
Particle identification is based on the \dEdx\ estimators \chipi\ (\chik ),
defined as the difference between the measured and expected ionization
expressed in terms of standard deviations for the $\pi$ ($\K$) mass hypothesis.
For charged tracks having momentum above $2~\GeVc$ a pion/kaon separation of
$2\sigma$ is achieved.  

\boldmath
\section{Measurement of $B(\decDW)/B(\decDR)$ }
\unboldmath


Starting from the sample of hadronic Z decays the $\DSTpm$ are
reconstructed as follows. Each pair of oppositely-charged tracks is
considered with the two mass assignments $\K^-\pi^+$ and $\pi^- \K^+$
and those with $\mid M(\K\pi) - M_{\D^0} \mid <30~\MeVcc$ are
retained. If both hypotheses satisfy the mass cut the event is
rejected.  In addition, the measured mean ionization of each track is
required to be closer, in terms of number of standard deviations, to
the expectation for the assumed mass hypothesis than for the
alternative hypothesis.  The decay angle $\theta^{\ast}_{\K}$ of the kaon in 
the $\D^0$ rest frame is required to satisfy $\mid \cos
\theta^{\ast}_{\K} \mid \leq 0.8$. Only combinations in which the two
tracks form a common vertex and each track has at least one VDET hit
are kept.

To build the $\DSTpm$ candidate the surviving track pairs are combined
with an extra charged track, the ``soft pion'' ($\pi_s$), of low
momentum, typically less than $4~\GeV/c$ ~\cite{ALEPH3} (the limits on
momentum are fixed by kinematics and resolution effects). In order to
reduce the combinatorial background, the energy of the $\DSTpm$
candidate is required to be greater than half the beam energy.

\begin{figure}[p]
  \begin{center}
    \leavevmode
    \setlength{\unitlength}{1.0mm}
    \begin{picture}(150,170)
      \put(20,0){\mbox{\psfig{figure=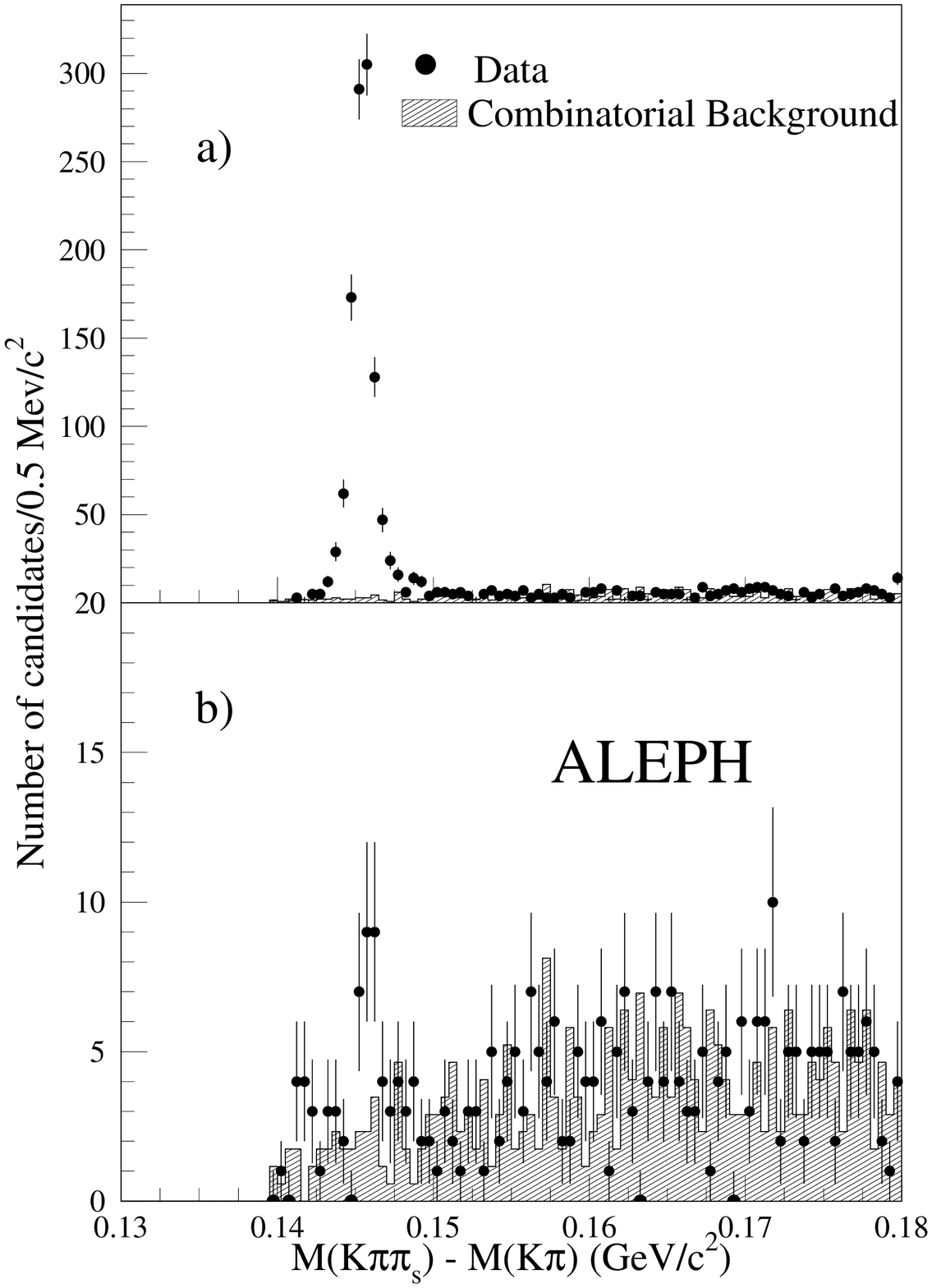,height=17cm}}}
    \end{picture}
    \caption[Mass-difference distribution]{Mass-difference distribution (a) for 
      candidates of the decay channel
      $\decDS$, $\decDR$ and (b) candidates of the decay channel $\decDS$, 
      $\decDW$. The dots with error bars are data while the hatched 
      histogram represents the distribution of the combinatorial background.}
    \label{fig:mass}
  \end{center}
\end{figure}

The Cabibbo-favoured decays of the $\D^0$ and $\bar{\D}^{0}$ are
contained in the sample for which the two pions have the same sign
(right-sign sample), while the DCSDs and mixing candidates are
contained in the sample for which the two pions have opposite sign
(wrong-sign sample).  The $\Delta M= M_{\pi_{s} \D^0} - M_{\D^0}$
distributions of the right-sign and wrong-sign samples are shown in
Fig.~\ref{fig:mass} together with the estimated combinatorial
background.  The shape of this combinatorial background is assumed to be 
the same as that obtained from events in the sideband region of the $\D^0$
invariant mass distribution above 2.1~$\GeVcc$, and is normalised to
the number of candidates having $\Delta M > 160~\MeVcc$.

Within a $\Delta M$ mass window from $143.5~\MeVcc$ to $147.5~\MeVcc$
the numbers of events in the right-sign and the wrong-sign samples
after the combinatorial background is subtracted are
$$ N_{\mathrm{RS}} = 1038.8 \pm 32.5(\stat) \pm 4.3(\syst) \ , $$
$$ N_{\mathrm{WS}} = 21.3 \pm 6.1(\stat) \pm 3.4(\syst) \ , $$
respectively. The systematic error is due to the limited statistics
used to determine the combinatorial background.

Monte Carlo studies show that, although the physics background
contamination in the right-sign sample is negligible, a small
contribution from physics backgrounds must be subtracted from the
wrong-sign sample. Four decay modes are found to contribute, namely:
$\D^0 \to \K^- \pi^+ (\pi^0)$, $ \D^0 \to \pi^- \mu^+ \nu_{\mu}$,
$\D^0 \to \pi^+ \pi^- \pi^0$ and $ \D^0 \to\K^- \K^+$.  They
contribute because of misidentification of one or both tracks or
because of missing neutrinos or $\pi^0$s.

\begin{figure}[htbp]
  \begin{center}
    \leavevmode
    \setlength{\unitlength}{1.0mm}
    \begin{picture}(150,170)
      \put(-15,0){\mbox{\psfig{figure=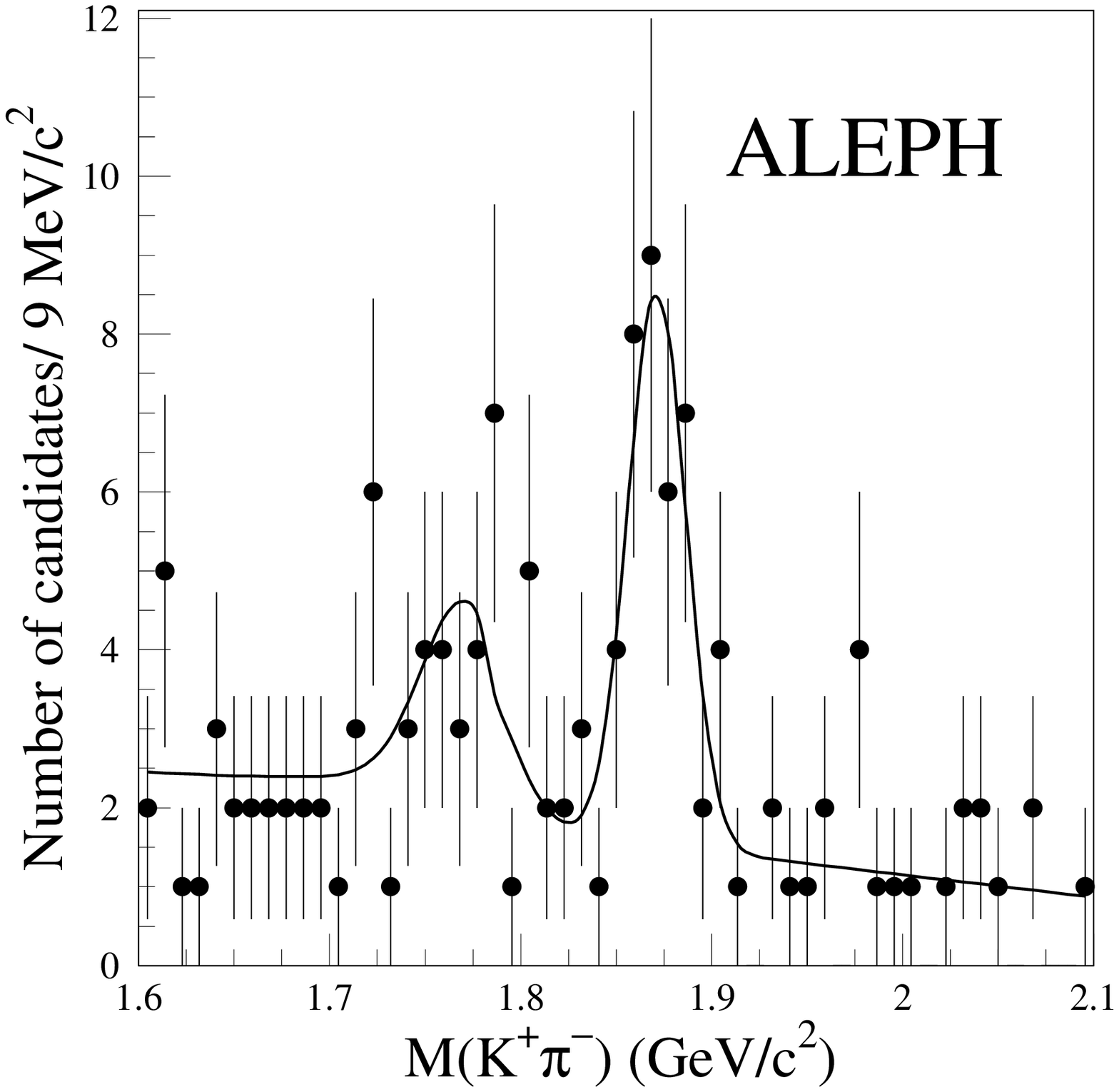,height=17cm}}}
    \end{picture}
    \caption[Mass distribution]{Mass distribution for  
      candidates of the decay channel $\decDS$, $\decDW$.
      A narrow peak at the nominal $\D^0$ mass ($M_{\D^0}=1864.5~\GeVcc$) 
      is present. The peak on the left is due to the decays 
      $\decKK$ and is fitted taking the shape from Monte Carlo.}
    \label{fig:d0mass}
  \end{center}
\end{figure}

To demonstrate that the peak appearing in Fig.~\ref{fig:mass}b is not related
to combinatorial background, events are selected in the $\Delta M$
mass window and the cut on $M(\K\pi)$ for the wrong-sign sample is not
applied. The resulting $M(\K\pi)$ distribution is shown in
Fig.~\ref{fig:d0mass}.  A narrow peak at the nominal $\D^0$ mass
($M_{\D^0}=1864.5~\GeVcc$) is present.  The peak on the left, due to
$\D^0 \to\K^- \K^+$ decays, is outside the $\D^0$ mass window.

Rather than relying on the Monte Carlo estimates for the physics
background subtraction, the data are used to estimate the contribution
of the dominant $\D^0 \to \K^- \pi^+ (\pi^0)$ physics background to
the wrong-sign sample.  This is achieved by repeating the selection,
with the $dE/dx$ requirement reversed, i.e. the ionization of the kaon
(pion) candidate has to be closer to the expectation for a pion
(kaon). This sample is hereafter called $\revc$.  In this sample the
$\D^0 \to \K^- \pi^+ (\pi^0)$ contribution is strongly enhanced while
the DCSD/mixing signal is suppressed by the same factor. The
contributions from the other decay channels remain the same, since the
cut is symmetric when the mass hypotheses are reversed.

According to Monte Carlo studies, the $\revc$ sample also contains
some small additional physics backgrounds, from semileptonic $\D^0$
decay channels, and these must also be taken into
account. Table~\ref{tab:phybkg} shows the number of expected physics
background candidates for the $\dirc$ and $\revc$ samples, calculated
from the Monte Carlo efficiencies and assuming the Particle Data Group
\cite{PDG96} branching ratios.

\renewcommand{\arraystretch}{1.1}
\begin{table}[htb]
\begin{center}
\leavevmode
\begin{tabular}{| l || c  | c  | }
\hline 
\multicolumn{1}{|c||}{} & {} & {}\\
\multicolumn{1}{|c||}{Channel} & {$\dirc$} & {$\revc$} \\
\hline \hline 
$\D^0 \to \K^- \pi^+ (\pi^0)$     & $ 1.56 \pm 1.08 $ & $26.48\pm 4.16$ \\
$\D^0 \to \pi^- \pi^+ \pi^0 $     & $ 0.36 \pm 0.36 $ & $0.36 \pm 0.36$ \\
$\D^0 \to \pi^- \mu^+ \nu_{\mu} $ & $ 0.16 \pm 0.16 $ & $0.16 \pm 0.16$ \\
$\D^0 \to \K^- \K^+            $  & $ 0.12 \pm 0.12 $ & $0.12 \pm 0.12$ \\
$\D^0 \to \K^- \e^+   \nu_{\e} $  &  $ - $            & $3.16 \pm 1.20$ \\
$\D^0 \to \K^- \mu^+  \nu_{\mu}$  &  $ - $            & $0.84 \pm 0.64$ \\
\hline

\end{tabular}
\caption[Physics background] {\small{Physics background estimated from the 
Monte Carlo with the standard and reversed $dE/dx$ cuts.}}
\label{tab:phybkg}
\end{center}
\end{table}

The expected contributions after the subtraction of the combinatorial
background to the number of candidates of the wrong-sign sample,
$N_{\mathrm{WS}}$, and the number of candidates of the $\revc$ sample,
$N_{\mathrm{WS}}^{\revc}$, can be written as
\begin{eqnarray}
 N_{\mathrm{WS}} & = & N_{\decDW} + N_{\K \pi (\pi^0)} + N_{\simm} \\
 N_{\mathrm{WS}}^{\revc} & = & \frac{1}{r} N_{\decDW} + r N_{\K \pi
 (\pi^0)} + N_{\simm} + N_{\rm other}
\label{sistema}
\end{eqnarray}
where the various quantities are explicited hereafter. 

\begin{itemize} 

\item {\boldmath$ N_{\decDW}$ } is the unknown number of events
attributed to the DCSD/mixing signal in the wrong-sign sample.
 
\item {\boldmath $N_{\K \pi (\pi^0)}$ } is the unknown number of
events attributed to the dominant $\D^0 \to \K^- \pi^+ (\pi^0)$
physics background in the wrong sign sample.

\item {\boldmath $r$} is the known enhancement factor for the $\D^0
\to \K^- \pi^+ (\pi^0)$ contribution obtained when the \dEdx\ cut is
reversed.  This is measured in the data to be $r=46.1 \pm 11.8$, by
applying the reversed \dEdx\ cuts to the right-sign sample and noting
the reduction in the size of the peak from the Cabibbo-favoured $\D^0$
decay.

\item {\boldmath $N_{\simm}$} is the estimate of the sum of the
backgrounds which are symmetric upon reversal of the \dEdx\ cut, 
i.e. the $\D^0 \to \pi^- \pi^+ \pi^0$, $\D^0 \to \pi^- \mu^+
\nu_{\mu}$ and $\D^0 \to \K^- \K^+ $ backgrounds.  $N_{\simm}$ is
assumed to be $r N_{\K \pi (\pi^0)} \cdot f_{\simm}$, where
$f_{\simm}$ is the fraction of symmetric events with respect the
number of $\K \pi (\pi^0)$ events in the Monte Carlo $\revc$ sample.
The error on this number is taken to be 100\% to take into account the
differences, compatible with statistical fluctuations, found both for the
$dE/dx$ and $\revc$ samples in Monte Carlo.
 
\item {\boldmath $N_{\rm other} $ }
is the additional backgrounds expected in the $\revc$ sample coming
from $\D^0 \to \K^- \e^+ \nu_{\e} $ and $\D^0 \to \K^- \mu^+
\nu_{\mu}$ decays. It is assumed to be $r N_{\K \pi (\pi^0)} \cdot
f_{\rm other}$, where $f_{\rm other}$ is also taken from the Monte
Carlo estimate. The uncertainties are due to the limited statistics
and to the errors on the branching ratios.

\end{itemize}

Using the values $N_{\mathrm{WS}} = 21.3 \pm 6.1(\stat) \pm 3.4(\syst)$,
$N_{\mathrm{WS}}^{\revc} = 56.4 \pm 8.1(\stat) \pm 2.3(\syst)$ measured in the data,
equations (1) and (2) yield
$$ N_{\K \pi (\pi^0)} = 1.03 \pm 0.15(\stat) \pm 0.27(\syst) \ ,$$
$$ N_{\decDW} = 19.1 \pm 6.1(\stat) \pm 3.5(\syst) \ .$$ 
The total physical background ($ N_{\K \pi (\pi^0)} + N_{\simm}$) is
$2.2 \pm 1.0$ and is consistent with the Monte Carlo expectation of
$2.2 \pm 1.2$ from Table~\ref{tab:phybkg}. 
The systematic uncertainties for $N_{\decDW}$ are listed in 
Table~\ref{tab:sysbr} and are derived by varying the following quantities 
by one standard deviation: (i) the systematic uncertainty 
on $N_{\mathrm{WS}}$ due to the combinatorial background subtraction, (ii)
the statistical and systematic uncertainties on the 
$N_{\mathrm{WS}}^{\revc}$ sample, (iii) the statistical uncertainty on 
$r$ and finally (iv) the uncertainties on the physics background processes 
as discussed previously.

Dividing $N_{\decDW}$ by the number of signal events
in the right-sign sample ($N_{\mathrm{RS}}$) yields a relative branching ratio of
$$ B(\decDW)/B(\decDR) = \left( 1.84 \pm 0.59(\stat) \pm 0.34(\syst) 
\right)\! \% \ .$$ 

\begin{table}[htb]
\begin{center}
\leavevmode
\begin{tabular}{|l|| c | c |}
\hline 
\multicolumn{1}{|c||}{Source} & Systematics \\ 
\hline \hline
Syst. error on $N_{\mathrm{WS}}$ & $\pm 3.4$ \\ 
Stat. error on $N_{\mathrm{WS}}^{\revc}$ & $\pm 0.3$ \\ 
Syst. error on $N_{\mathrm{WS}}^{\revc}$ & $\pm 0.1$ \\ 
$r =$ effic.($dE/dx$)/effic.($\revc)$ & $\pm 0.3$ \\ 
Fractions of physics background & $\pm 0.9$ \\ 
\hline Total & $\pm 3.5$ \\ 
\hline
\end{tabular}
\caption[Systematics]{\small{Systematic uncertainties for the
measurement of the number of $\decDW$ decays.}}
\label{tab:sysbr}
\end{center}
\end{table}

\section{Proper Time Distribution}

Assuming small mixing and neglecting $\mathrm{CP}$-violating terms,
the time evolution for the signal in the wrong sign sample is expected
to have the following form~\cite{formula}
\begin{equation}
N_{\decDW}(t) \propto \left[ R_{\DCSD} + \sqrt{ 2 R_{\mx} R_{\DCSD} }
t/\tau \cos \phi + R_{\mx} \frac{1}{2} (t/\tau)^2 \right] e^{-t/\tau}
\ ,
\label{temppf} 
\end{equation}
where $R_{\DCSD}$ is the ratio of doubly Cabibbo-suppressed over
Cabibbo-favoured decays, $R_{\mx}$ is the ratio $B(\D^0 \to
\bar{\D}^0 \to \K^+ \pi^-)/B(\D^0 \to \K^- \pi^+)$ and $\cos \phi$ is
the phase angle parametrizing the interference between the two
processes.  The first term, due to the DCSD decay, has the
conventional exponential proper time dependence with a decay constant
given by the $\D^0$ lifetime.  The third term is the
contribution from the mixed events and peaks at $t/\tau=2$.  The
second term accounts for possible interference between both processes.
The significant differences in the structure of the proper time
distributions for the DCSD and the mixing signals allow their
respective contributions to be estimated.

The proper time $t = \ell \frac{M}{p}$
of a $\D^0$ candidate is calculated from the decay length 
$\ell$, defined as the distance between the primary vertex and the $\D^0$ 
decay vertex projected along the direction of flight of the $\D^0$, and the
reconstructed momentum $p$ and mass $M$ of the candidate. 
The average resolution on the proper time is $\approx 0.1$~ps
and is dominated by the uncertainty on the position of the $\D^0$ vertex. 

\begin{figure}[htbp]
  \begin{center}
    \leavevmode
    \setlength{\unitlength}{1.0mm}
    \begin{picture}(150,170)
      \put(20,0){\mbox{\psfig{figure=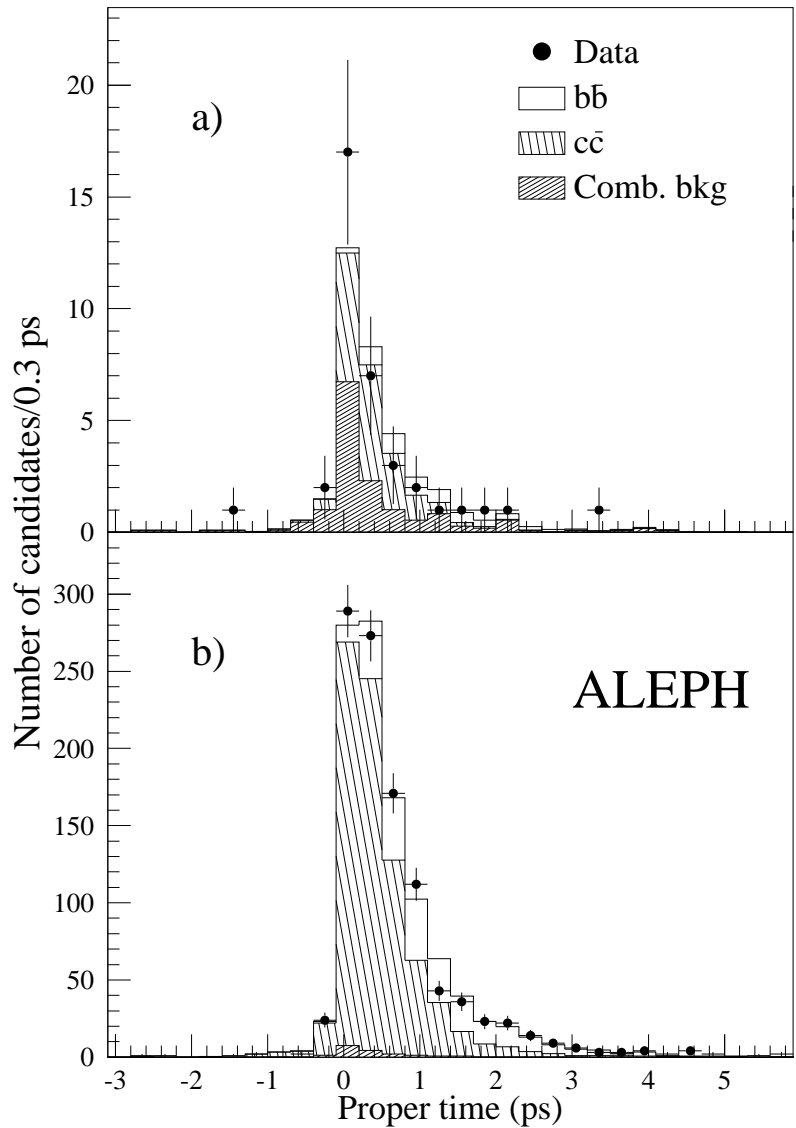,height=17cm}}}
    \end{picture}
    \caption[Proper time distribution]
     {Proper time distribution for (a) the $\decDW$ candidates, and
     (b) for the $\decDR$ candidates. The dots with error bars are the data. 
     The histograms are the contributions of $\cc$, $\bb$ and combinatorial 
     background events resulting from the unconstrained fit when no interference
     is assumed.}
    \label{fig:proptime}
  \end{center}
\end{figure}

The distributions of proper time in the signal region for the wrong-sign 
sample observed in the data is shown in Figure~\ref{fig:proptime}a.
The result of a binned maximum likelihood fit is also shown. 
The following contributions are included in the probability density
of the likelihood function:

\begin{itemize} 
\item A direct $\cc \to \D^0 \X$ component, which has the 
proper time dependence given by Eq.~\ref{temppf} convoluted with the 
detector resolution. The number of events attributed to the 
DCSD signal and the mixing signal are left free in the fit. 
Various assumptions for the phase of the interference term are investigated. 

The fraction of the signal which is attributed to $\cc \to \D^0 \X$, 
(rather than the $\bb \to \cc \to \D^0 \X$ discussed next)
is $f_{\crm}^{\K\pi} = \left(77.0 \pm 2.7(\stat) \pm 0.5(\syst)\right)\! \%$.  
It is extracted from the data using a likelihood fit to the proper time 
dependence of the right-sign sample. 
This fit is essentially the same as the fit to the wrong-sign sample except that 
proper time dependence of the Cabibbo-favoured events is  
assumed to be exponential, as expected for small mixing, and the contribution accounting 
for the physics background is not necessary. The result of this fit is shown 
in Figure~\ref{fig:proptime}b. 

\item An indirect $\bb \to \cc \to \D^0 \X$ component, which
has the same proper time dependence as the direct component of the signal, 
but modified to take into account the effect of the additional flight 
distance of the B meson. The exact shape for this component is taken from 
Monte Carlo after appropriate reweighting for the world average B hadron and $\D^0$ 
lifetimes~\cite{PDG96}. 

\item A physics background contribution for which the proper time dependence is assumed 
to be exponential with a decay constant given by the $\D^0$ lifetime. The number of 
events attributed to this process is $2.2 \pm 1.0$, as determined in section 3.

\item A combinatorial background contribution, for which the proper
time dependence is measured in the data from the sideband regions of
the $\Delta M$ plot.  The number of events attributed to this process
is $15.7 \pm 3.4$ events.  

\end{itemize}

For all the above contributions, except the combinatorial background,
an additional proper time smearing of $(29 \pm 12)\%$ is applied to
take into account that the proper time resolution measured in the data
is slightly worse than that predicted by the Monte Carlo. This Monte
Carlo/data comparison is performed by selecting candidates from the
side band of the $\K\pi$ mass distribution both in Monte Carlo and
data and comparing the width of a Gaussian fit to the negative proper
time distribution of these events.  To enhance the fraction of tracks
in this sample coming from the interaction point the contamination
from $\cc$ and $\bb$ events is suppressed by applying to the opposite
hemisphere the lifetime tag veto described in Ref.~\cite{Rb}.

Setting the interference term to zero ($\cos \phi =0$), the result of
the fit to the proper time distribution of the wrong-sign sample is
\begin{eqnarray*}
N_{\DCSD} & = & 20.8 {}^{+8.4}_{-7.4}(\stat) \pm 4.0(\syst) \ , \\
N_{\mx} & = & -2.0 \pm 4.4(\stat) \pm 1.1(\syst) \ .
\end{eqnarray*}
The fitted value for $N_{\mx}$ is outside the physical region, thus no
mixing is observed.  Table~\ref{tab:fit} summarises all the sources of
systematic uncertainty considered. They are computed by varying by one
standard deviation on the fit the additional proper time smearing, the
$\D^0$ lifetime, the charm fraction and the combinatorial and 
physical background rates.
The systematic uncertainties due to the combinatorial background
proper time shape is evaluated by repeating the fit many times with a
new combinatorial background shape, obtained by randomly varying the
contents of each proper time bin according to a Poisson distribution.

\begin{table}[tb]
\begin{center}
\leavevmode
\begin{tabular}{|c|l  r|  r|}
\hline
                         &  \multicolumn{2}{c |}{unconstrained fit}  &  \multicolumn{1}{c |}{constrained fit}      \\
\cline{2-4}
Source  of uncertainty   &  $N_{\mx}$      &   $N_{\DCSD}$  & $N_{\DCSD}$               \\
\hline \hline                                                                          
Resolution               &  $ < 0.1$       & $ \pm 0.1  $   & $ < 0.1$       \\
$\D^0$ lifetime          &  $ < 0.1$       & $ < 0.1 $      & $ < 0.1$       \\ 
Comb. back. distribution & $\pm 0.7$       & $ \pm 1.4$     & $ \pm 1.1$     \\
Comb. back. rate         & $\pm 0.8$       & $ \pm 3.6 $    & $\pm 2.8$      \\
Charm fraction           & $\pm 0.4$       & $ \pm 0.3$     & $ \pm 0.1 $    \\
Phys. back. rate         &                 & $ \pm 1.0 $    & $  \pm 1.0 $   \\
\hline
Total                    & $\pm 1.1$       &   $\pm 4.0$    &             $ \pm 3.2$       \\
\hline 
\end{tabular}

\caption[Syste2]{\small {List of systematic uncertainties contributing 
to $N_{\DCSD}$ and $N_{\mx}$ for the case of no interference. The constrained 
fit results are obtained with $N_{\mx}$ constrained to be non-negative.}} 
\label{tab:fit}
\end{center}
\end{table}

\begin{figure}[htbp]
  \begin{center}
    \leavevmode
    \setlength{\unitlength}{1.0mm}
    \begin{picture}(150,170)
      \put(-5,0){\mbox{\psfig{figure=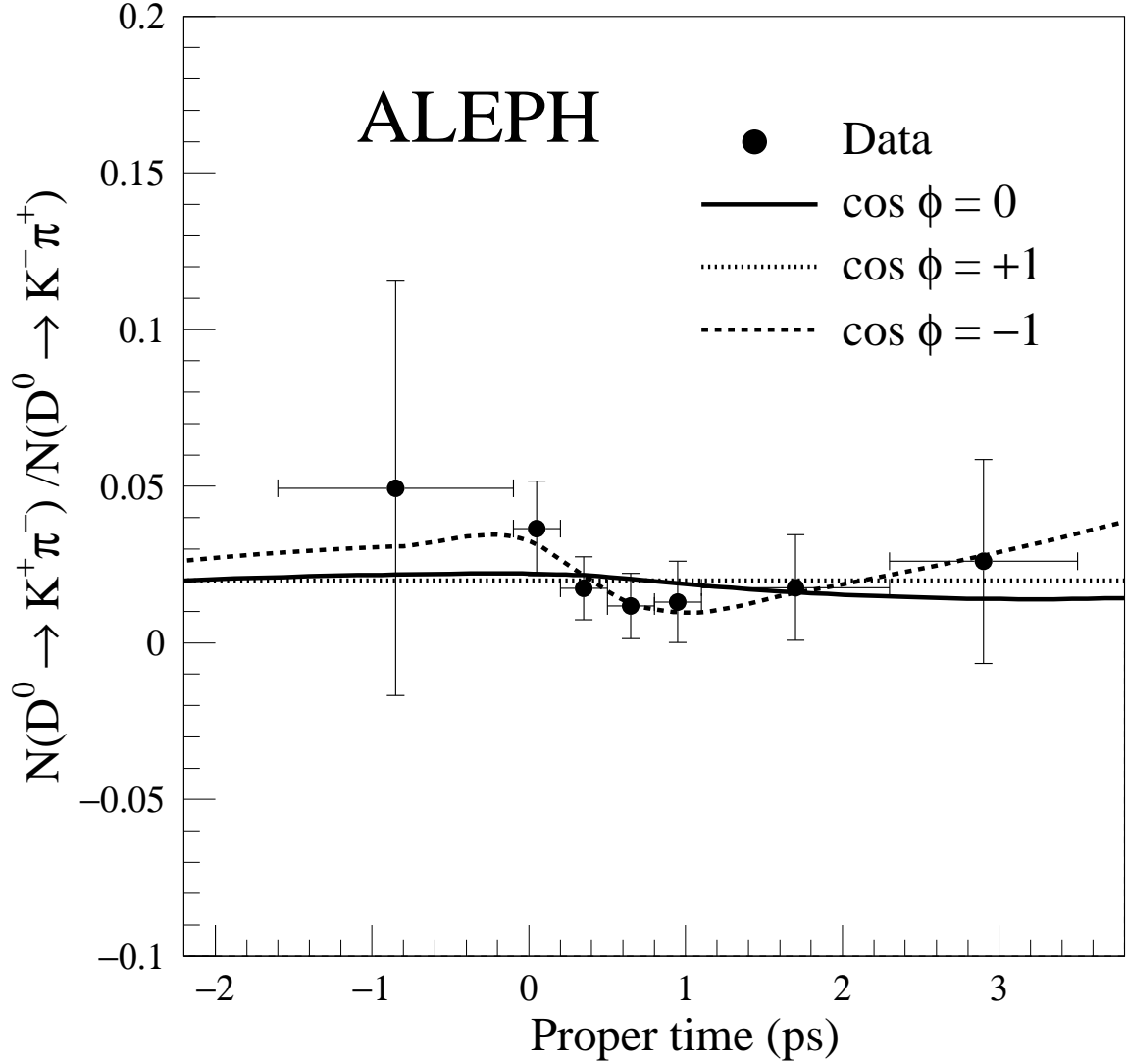,height=17cm}}}
    \end{picture}
     \caption[time ratio]
      {Proper time distribution of the $\decDW$ over $\decDR$ 
      candidates with the background subtracted. The dots with error bars 
      are data. The errors are the sum in quadrature of statistical
      and systematic uncertainties due to combinatorial background 
      subtraction. The solid curve is the fit result with no interference, the dotted
      and dashed curves are the fit results assuming fully 
      constructive and destructive interference, respectively.}
    \label{fig:mixsub}
  \end{center}
\end{figure}

Figure~\ref{fig:mixsub} shows the time 
dependence of the ratio of wrong-sign over the right-sign candidates 
after the background subtraction.

\begin{figure}[htbp]
  \begin{center}
    \leavevmode
    \setlength{\unitlength}{1.0mm}
    \begin{picture}(150,170)
      \put(-5,0){\mbox{\psfig{figure=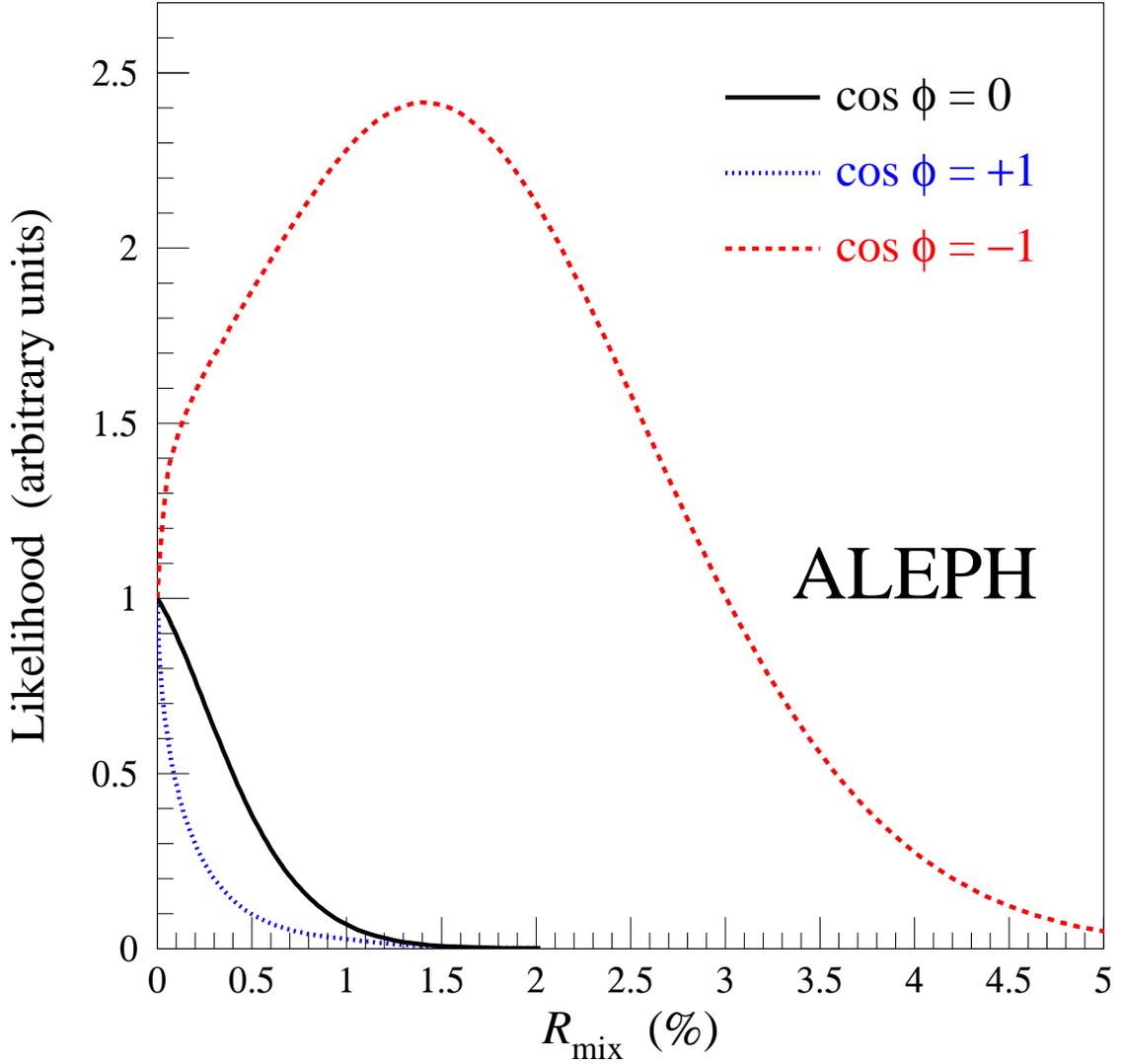,height=17cm}}}
    \end{picture}
    \caption[Likelihood]
     {Likelihood of the fit as a function of $R_{\mx}$ calculated by
     leaving free the $R_{\DCSD}$ parameter and constraining all
     the external parameters within their uncertainties in 
     order to take the systematic errors into account.
     The solid line represents the fit results with no interference, the 
     dotted line is the likelihood for fully positive interference and the 
     dashed line for fully negative interference.}
    \label{fig:like}
  \end{center}
\end{figure}

If $N_{\mx}$ is constrained to be non-negative, the result is
$$ N_{\DCSD} = 18.4 {}^{+6.2}_{-5.8}(\stat) \pm 3.2(\syst) \ ,$$
$$ N_{\mx} = 0^{+3.0}_{}(\stat) {}^{+1.4}_{}(\syst) $$ which
translates to
$$ R_{\DCSD} = \left( 1.77 {}^{+0.60}_{-0.56}(\stat) \pm 0.31 (\syst)
\right) \! \% \ .$$ 
In this case the systematic uncertainty on
$N_{\mx}$ is obtained by adding to the likelihood used to fit the
data, additional Gaussian terms for the extra smearing on the time resolution, 
the $\D^0$ lifetime, the charm fraction and the physical background rate,
and 
Poissonian terms for the combinatorial background rate and
shape. Figure~\ref{fig:like} shows the resulting likelihood as a
function of the assumed mixed fraction.  By integrating the resulting
likelihood over the allowed region an upper limit of $N_{\mx} < 9.6$
at 95\% confidence level is obtained, corresponding to
$$ R_{\mx} < 0.92 \% \ \ {\mathrm{at}} \ 95\% \ {\mathrm{CL.}} $$

The effect of interference has been studied by fitting the data with fully 
constructive ($\cos \phi = +1$) and fully destructive interference 
($\cos \phi = -1$); the respective upper limits are  
$R_{\mx} < 0.96 \% \ \ {\mathrm{at}} \ 95\% \ {\mathrm{CL}}$
and
$R_{\mx} < 3.6 \% \ \ {\mathrm{at}} \ 95\% \ {\mathrm{CL}}$.
For the case 
$\cos \phi = -1$ $N_{\mx} = 14.8^{+12.1}_{-13.3}(\stat) \pm 3.3(\syst)$.

\section{Conclusion}

The $\decDW$ decay is studied to determine the branching ratio 
$B(\decDW)/B(\decDR)$. The method consists in reconstructing the $\decDS$ 
decay where the $\D^0$ can subsequently decay to $\K^+ \pi^-$ or $\K^- \pi^+$.
The numbers of reconstructed decays observed after background subtraction give 
$$ B(\decDW)/B(\decDR) = \left(1.84 \pm 0.59(\stat) \pm 0.34(\syst) 
\right)\! \% \ . $$ 
This is 1.4 standard deviations from the CLEO~\cite{CLEO} measurement
$B(\decDW)/B(\decDR) = \left(0.77 \pm 0.25 (\stat) \pm 0.25(\syst) \right)\!\%$.

In order to distinguish the two possible contributing processes, the fraction of
doubly Cabibbo-suppressed decays $R_{\DCSD}$ and $\D^0$--$\bar{\D}^0$ mixing rate $R_{\mx}$, the 
proper time distribution is analysed, yielding
$$ R_{\DCSD} = \left( 1.77 {}^{+0.60}_{-0.56}(\stat) \pm 0.31 (\syst) 
\right) \! \% \ , $$
$$ R_{\mx} < 0.92 \% \ \ {\mathrm{at}} \ 95\% \ {\mathrm{CL}} \ ,$$
assuming no interference between the two processes.
The fit is improved if destructive interference is allowed.

This can be compared with the results obtained by the E691
Collaboration~\cite{Anjos}: $R_{\DCSD} < 1.5 \%$ at 90\% CL 
based on the number of observed events $N_{\DCSD} = 1.8 \pm 13.2$, and $R_{\mx} < 0.5 \%$ at
90\% CL.  The E791 collab.~\cite{E791} finds $R_{\DCSD}=
\left(0.68{}^{+0.34}_{-0.33}(\stat) \pm0.07(\syst)\right)\%$, for $R_{\mx} = 0$, 
in agreement within 1.5 standard deviations with our result, and sets a
limit $ R_{\mx} < 0.85 \% \ \ {\mathrm{at}} \ 90\% \ {\mathrm{CL}}$ allowing 
CP violation in the interference term.

\section*{Acknowledgement}

We wish to congratulate our colleagues in the CERN accelerator
divisions for successfully operating the LEP storage ring. We are
grateful to the engineers and technicians in all our institutions for
their contribution towards ALEPH's success. Those of us from
non-member countries thank CERN for its hospitality.

\newpage

\end{document}